\documentclass[sigconf]{acmart}

\usepackage{seqsplit}
\usepackage{url}
\usepackage{listings}
\usepackage{color}
\usepackage{xcolor}
\usepackage{textcomp}

\colorlet{punct}{red!60!black}
\definecolor{background}{HTML}{EEEEEE}
\definecolor{delim}{RGB}{20,105,176}
\colorlet{numb}{magenta!60!black}

\lstdefinelanguage{json}{
    basicstyle=\scriptsize,
    numbers=left,
    numberstyle=\scriptsize,
    stepnumber=1,
    numbersep=8pt,
    showstringspaces=false,
    breaklines=true,
    frame=single,
    literate=
     *{0}{{{\color{numb}0}}}{1}
      {1}{{{\color{numb}1}}}{1}
      {2}{{{\color{numb}2}}}{1}
      {3}{{{\color{numb}3}}}{1}
      {4}{{{\color{numb}4}}}{1}
      {5}{{{\color{numb}5}}}{1}
      {6}{{{\color{numb}6}}}{1}
      {7}{{{\color{numb}7}}}{1}
      {8}{{{\color{numb}8}}}{1}
      {9}{{{\color{numb}9}}}{1}
      {:}{{{\color{punct}{:}}}}{1}
      {,}{{{\color{punct}{,}}}}{1}
      {\{}{{{\color{delim}{\{}}}}{1}
      {\}}{{{\color{delim}{\}}}}}{1}
      {[}{{{\color{delim}{[}}}}{1}
      {]}{{{\color{delim}{]}}}}{1},
}

\definecolor{listinggray}{gray}{0.9}
\definecolor{lbcolor}{rgb}{0.9,0.9,0.9}
\definecolor{lightgray}{rgb}{.9,.9,.9}
\definecolor{darkgray}{rgb}{.4,.4,.4}
\definecolor{purple}{rgb}{0.65, 0.12, 0.82}
\definecolor{forestgreen}{rgb}{0.13, 0.55, 0.13}

\lstdefinelanguage{JavaScript}{
  keywords={typeof, new, true, false, catch, function, return, null, catch, switch, var, if, in, while, do, else, case, break},
  keywordstyle=\color{blue}\bfseries,
  ndkeywords={class, export, boolean, throw, implements, import},
  ndkeywordstyle=\color{black}\bfseries,
  identifierstyle=\color{black},
  sensitive=false,
  comment=[l]{//},
  morecomment=[s]{/*}{*/},
  commentstyle=\color{forestgreen}\ttfamily,
  stringstyle=\color{purple}\ttfamily,
  morestring=[b]',
  morestring=[b]"
}
\newcommand{\code}[1]{{\texttt{\small#1}}}

\newboolean{showcomments}
\setboolean{showcomments}{true}
\ifthenelse{\boolean{showcomments}}
{\newcommand{\nb}[2]{
\fbox{\bfseries\sffamily\scriptsize#1}
{\sf\small$\blacktriangleright$\textit{#2}$\blacktriangleleft$}
}
}
{\newcommand{\nb}[2]{}
}

\AtBeginDocument{%
  }

\copyrightyear{2025}
\acmYear{2025}

\acmConference[]{arXiv preprint}{Feb 2025}{arXiv}

\begin{document}

\title{Browser Fingerprint Detection and Anti-Tracking}



\author{Kaitong Lin}
\affiliation{%
  \institution{New York Institute of Technology}
  \city{Vancouver}
  \country{Canada}}
\email{klin16@nyit.edu}

\author{Huazhu Cao}
\affiliation{%
  \institution{New York Institute of Technology}
  \city{Vancouver}
  \country{Canada}}
\email{hcao06@nyit.edu}

\author{Amin Milani Fard}
\affiliation{%
  \institution{New York Institute of Technology}
  \city{Vancouver}
  \country{Canada}}
\email{amilanif@nyit.edu}


\begin{abstract}
Digital fingerprints have brought great convenience and benefits to many online businesses. However, they pose a significant threat to the privacy and security of ordinary users. In this paper, we investigate the effectiveness of current antitracking methods against digital fingerprints and design a browser extension that can effectively resist digital fingerprints and record the website's collection of digital fingerprint-related information.

\end{abstract}




\maketitle

\section{Introduction}

Digital fingerprints of a browser are collections of various user information, such as browser and system version and device attributes, which are relatively fixed and difficult to avoid or delete. They are widely used in the Internet industry for purposes such as identity verification on social networks, advertising platforms, and financial services. These fingerprints enable accurate identification of individuals across websites and devices, often without consent, posing a significant threat to privacy. The harm that invasive tracking brings is greater than that of cookies. Even if users have some anti-tracking awareness and use some anti-tracking measures, such as regularly cleaning up browser local storage or using private browsing mode, they may still be tracked \cite{vastel2018fp}. Although many browser manufacturers have pointed out the dangers of digital fingerprint tracking, the use of digital fingerprints has increased significantly over the past decade \cite{mozillaSecurityAntiTracking}. Preventing the misuse of digital fingerprints is critical to ensuring higher privacy protection.

Digital fingerprint algorithms often use fuzzy hashing technology, which maps similar feature sets to similar or identical fingerprint values. For example, a slight change in the browser's screen resolution or time zone will not significantly affect the final hash value because the algorithm focuses on patterns of overall features rather than single details. Therefore, changing the digital fingerprint requires many feature changes, which is difficult for ordinary users. Hence, an anti-digital fingerprint extension that detects and disguises identities is significant for ordinary users. It needs to help users reduce the risk of digital fingerprint tracking in an easy-to-understand and easy-to-use way. 

\textbf{Contributions.} Existing approaches have limitations in mitigating browser fingerprint tracking. In this work, we propose a solution by developing a user-friendly and effective Chrome extension, available for download\footnote{\url{https://github.com/nyit-vancouver/browser-fingerprint-detector-and-antitracking-extension}}, with the following core features:

\begin{itemize}
    \item \textbf{Fingerprint Detection}: The extension will detect and analyze the users' current browser fingerprint, allowing them to have an overview and verify validity after spoofing the fingerprint. We use JavaScript APIs and objects that provide device information to implement this.
    \item \textbf{Fingerprint Customization \& Randomization}: Users can customize spoofed browser fingerprints by selecting the provided configuration or toggling a one-click switch. The solution rewrites APIs and objects for accessing fingerprint data and modifies their return values.
    \item \textbf{Tracking Monitoring}: It monitors and logs websites that attempt to track the user using the browser fingerprint. Requests are intercepted to access browser fingerprint information, and logging is recorded when data is accessed.
    \item \textbf{Tracking Dashboard}: Based on the logs generated by the Tracking Monitoring feature, it offers users an overview of the number of times fingerprints have been accessed. To make it accessible and more user-friendly, it uses a chart library to display statistics.
\end{itemize}

\section{Related Work}

Current precautions to prevent Internet tracking include:

\textbf{Private/Incognito Mode.} Enabling the privacy mode of the browser will limit the saving of history records and block the impact of cookies and cached data. Although incognito mode can protect against some simple tracking technologies, it is almost ineffective in defending against digital fingerprints. This is because digital fingerprints use much information, including, but not limited to, browser and hardware device information. Many attributes will not change because the user uses privacy mode. Therefore, privacy mode cannot effectively prevent digital fingerprint tracking \cite{wu2017evaluating}.

\textbf{Ad Blockers \& Privacy Badger.} Ad blockers and Privacy Badger reduce tracking by blocking ads and certain tracking scripts. They are often very effective against common third-party tracking. However, digital fingerprints are often generated by legitimate browser behavior, so these extensions cannot effectively identify and block this situation \cite{mughees2017detecting}.

\textbf{Disabled JavaScript.} Enabling the browser's JavaScript disabling function can effectively prevent device and browser information leakage caused by certain tracking scripts. However, this behavior will seriously affect the typical web browsing experience. In addition, digital fingerprint technology can also collect some user information through HTTP header information \cite{bortolameotti2020headprint}.

\textbf{Tor Browser.} The randomization and standardized browser settings of Tor Browser can significantly reduce the uniqueness of fingerprints and prevent users from being uniquely identified. Although Tor is effective in preventing digital fingerprint tracking, its performance and speed are much lower than those of other conventional browsers. The use of Tor is also tricky and requires some relevant knowledge, making it less effective in terms of user experience and unsuitable for daily use by the public \cite{bortolameotti2020headprint}. 

\textbf{Chameleon.} Chameleon is a browser extension that supports Firefox. It mainly provides functions such as confusing fingerprints through custom options and predefined modes. However, it focuses more on modifying information that is not significantly represented in digital fingerprints. In addition, the configuration is global. Users cannot set a configuration on certain web pages or know what information the website monitors. Meanwhile, users still need some relevant technical knowledge before using it.

\section{The Proposed Solution}

\begin{figure}[h]
\centering
\includegraphics[trim = 6mm 2mm 6mm 0mm,width=1.0\hsize]{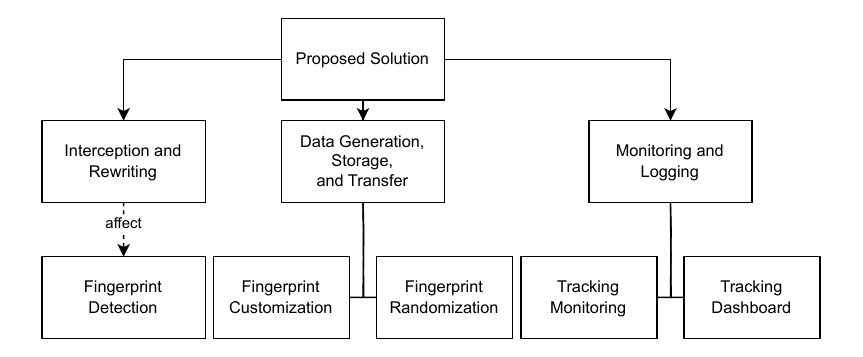}
\caption{The proposed system components}
\label{fig:components}
\end{figure}

\begin{figure*}[h]
\centering
\includegraphics[trim = 10mm 35mm 10mm 10mm,width=0.6\hsize]{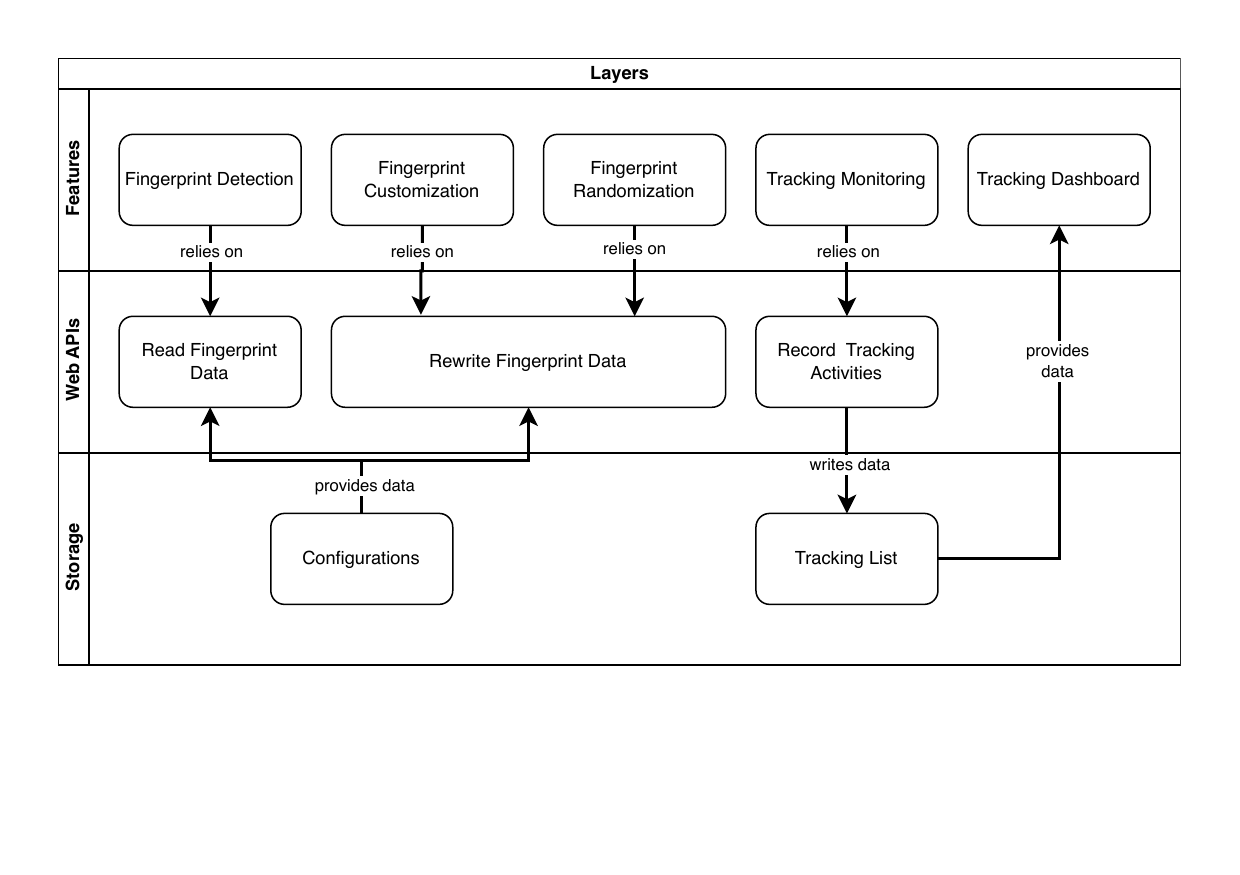}
\caption{Architectural diagram of our browser fingerprint detection and anti-tracking.}
\label{fig:1}
\end{figure*}

To reduce the risk of browser fingerprint tracking for regular users, we propose a solution by developing a user-friendly and effective Chrome extension since Google Chrome plays a dominant role in the browser market \cite{chromeRole}. Figure \ref{fig:components} illustrates our proposed system that is divided into three main components: (1) Rewriting and Interception Module, (2) Data Generation, Storage and Transfer Module, and (3) Monitoring and Logging Module. The architectural diagram of our design is shown in Figure \ref{fig:1}.

\subsection{Rewriting and Interception Module}
This module rewrites critical APIs and the request headers related to browser fingerprinting. Generally, a website collects browser fingerprints in two main ways \cite{collectFingerprint}: (1) by calling specific Javascript APIs, and (2) by obtaining certain request headers from a request sent to the Web server.

\subsubsection{API Rewriting.} JavaScript is a widely used browser programming language \cite{js} that provides access to the browser environment through the global object \code{window}. The \code{window} object contains a large number of properties and methods \cite{jsWindow}, such as \code{navigator}, \code{screen}, and \code{document}, which are often used for fingerprint analysis of the browser. The function \code{Object.defineProperty} allows us to define a new property on an object or modify an existing one \cite{defineProperty}. With this function, we can modify the properties and methods related to browser fingerprinting \cite{definePropertyUsage}, even if some are read-only. For example, with the following code, we can modify the value of \code{navigator.userAgent} and get a forged value when calling it:
\begin{lstlisting}
const originalValue = window.navigator.userAgent

Object.defineProperty(window.navigator, 'userAgent', {
  get() {
    return 'Forged UserAgent' // return a forged value here
  },
  configurable: true
})

window.navigator.userAgent // 'Forged UserAgent'
\end{lstlisting}

\subsubsection{Request Interception}

Browser fingerprints can be accessed from request headers sent to the server. When a client requests resources, headers are included in the request. Headers such as \code{User-Agent} and \code{Accept-Language} can be used to identify information such as the user's browser, operating system, and personalized settings, which can be related to user identification. There are functions related to web requests in Chrome extension development, such as \code{\seqsplit{chrome.webRequest.onBeforeSendHeaders}} \cite{onBeforeSendHeaders} and \code{\seqsplit{chrome.declarativeNetRequest}} \cite{declarativeNetRequest}, which can be utilized in the solution.

\subsection{Data Generation, Storage and Transfer Module}
This module generates forged fingerprint data based on the user configuration. The solution provides configurable options for key browser characteristics as \code{language} and \code{userAgent}. For dynamic information such as \code{Canvas}, \code{WebGL}, and \code{AudioContext}, it introduces random noise into these fingerprint values. This approach effectively alters the fingerprint while maintaining the visual and aural integrity of the rendered content.

\subsubsection{Data Generation}
With the one-click antitracking feature enabled, the system selectively modifies the key browser characteristic in the following ways:

1. Select specific device information, including but not limited to \code{Canvas},  \code{WebGL}, and \code{AudioContext}. As it is more representative, websites tend to track this information to generate browser fingerprints.

2. For static information such as \code{userAgent} and \code{language}, randomly select values in a predefined configuration list. For dynamic information such as \code{WebGL} and \code{Canvas}, inject random noise into sensitive rendering processes.

This strategy ensures that the generated fingerprints look real and varied, thus effectively enhancing privacy protection without compromising the user experience. After being generated and configured, the data will be stored based on the current tab ID, which allows the forged digital fingerprints to remain consistent across a single browsing session while being configuration-independent across multiple browsing sessions.

\subsubsection{Data Storage and Clearing Mechanism}
Compared to Web Storage, Chrome extension development provides more suitable storage. In Chrome extension development, Chrome Storage API provides four different data storage approaches \cite{chromeStorage4api}. For our proposed solution, two types of data should be stored, and we utilize \code{chrome.storage.session} and \code{chrome.storage.local} based on their respective strengths in handling different data types.

1. \textit{Configuration Data}: Since configuration data serves only the current webpage and has a life cycle tied to the webpage duration, it is stored temporarily in \code{chrome.storage.session}. This session storage, with a capacity of up to 10 MB (1 MB in Chrome versions \( \leq \)111) \cite{sessionStorage}, ensures that data is automatically cleared when the browser is closed. When the current tab is closed, our extension actively clears the data to preserve normal space in memory.

2. \textit{Log Data}: It requires long-term persistence to provide detailed insight for the \textit{Tracking Dashboard} feature over time. To meet this requirement, we store the logs in \code{chrome.storage.local}, allowing a maximum capacity of 10 MB (5 MB in Chrome versions \( \leq \)113) \cite{localStorage}. To optimize space, our solution clears 20\% of the storage when the remaining capacity drops below 10\%, ensuring that the storage does not overflow without deleting too much data.

\subsubsection{Data Transfer Process}
In the proposed solution, we differentiate two different environments:

1. \textit{Chrome Extension Environment}: In this environment, Chrome \code{storage} and \code{declarativeNetRequest} APIs can be called to store data and intercept requests, and tab-related event listeners can be set.

2. \textit{Web Environment}: It allows for JavaScript API rewriting and interactions within the webpage.
Communication between these two environments is facilitated by the JavaScript \code{CustomEvent} interface \cite{customEvent}, which can send custom data.

There are two primary components in Chrome extension development. (1) The background script is a service worker script that runs independently of web pages \cite{background}. It listens for browser events, allowing monitoring and reaction to events in the browser. It is loaded when the extension is launched and unloaded when disabled or uninstalled. (2) The content script runs within the context of the web page \cite{content}. By configuring 'matches', 'run\_at', and 'world' in \code{manifest.json}, the content script will be injected into webpage JavaScript when webpages with any URLs are loading:

\lstset{
  literate={<}{{<}}1 {>}{{>}}1 {_}{{\_}}1, 
  basicstyle=\ttfamily\scriptsize,
  breaklines=true,                        
  mathescape=false                         
}

\begin{lstlisting}[language=JavaScript, mathescape=false, caption={manifest.json}]
// manifest.json
{
  "content_scripts": [
    {
      "matches": ["<all\_urls>"],
      "js": ["./static/js/content.js"],
      "run_at": "document_start",
      "world": "MAIN" // share the same execution environment as the webpage's JavaScript
    }
  ]
}
\end{lstlisting}

Since the content script in our proposed solution can only call functions in the webpage, it can be categorized to the web environment.

\begin{figure*}[t]
\centering
\includegraphics[trim = 10mm 15mm 15mm 10mm,width=0.58\hsize]{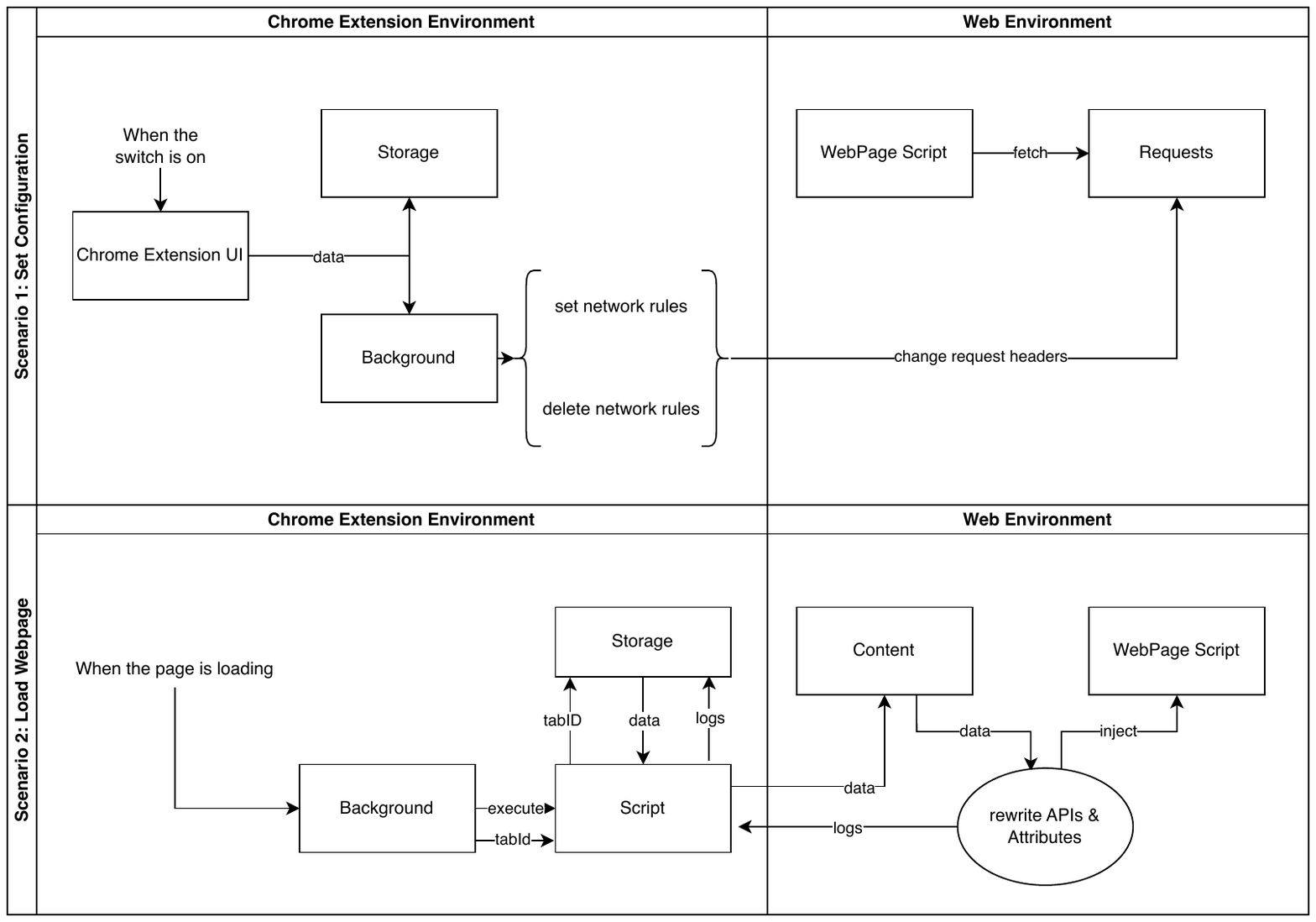}
\caption{Architectural diagram of two scenarios between two environments.}
\label{fig:twoScenarios}
\end{figure*} 

As shown in Figure \ref{fig:twoScenarios}, in general, there are two scenarios:

\textbf{Scenario 1: Set Configuration.} When a configuration is set, such as turning on the switch or customizing settings, configuration data will be stored in \code{chrome.storage.session}. If the configuration affects request headers, the Chrome extension's user interaction (UI) sends data to the background script via \code{chrome.runtime.sendMessage}. The background script listens for these messages and applies configurations using \code{chrome.declarativeNetRequest} to modify request headers accordingly.

\textbf{Scenario 2: Load a Webpage.} When a webpage is loaded or refreshed, the background script detects the event and executes the Script function, passing the tab ID. The Script function retrieves the configuration data for the tab from \code{chrome.storage.session}. If configuration data exists, it dispatches a \code{CustomEvent} with the configuration details. The Script function listens for log events from the content script.

Simultaneously, the content script is injected into the webpage and listens for custom events from the background script. When receiving a configuration event, JavaScript APIs will be rewritten, the logging logic will be integrated, and configuration will be applied to the modified API responses. The content script injection into the webpage is temporary, and the rewriting APIs before refreshing will be set to default after another refresh because it is not part of the webpage resources. Hence, it is necessary to rewrite them after each refresh. 
If a rewritten API is called, the \textit{Log Collection} Module generates batch log data and sends it to the script via a \code{CustomEvent} to write the log data asynchronously into \code{chrome.storage.local}.

\subsection{Monitoring and Logging Module}
This module logs the number of rewritten API calls within a visit of one page. These logs are implemented in the \textit{Intercept and Rewrite} Module, and the logs are stored to storage asynchronously without blocking subsequent code execution. This provides insight into how often a site attempts to collect fingerprinting data while minimizing the impact on page performance, allowing users to have a more intuitive understanding of browser fingerprinting and evaluate the effectiveness of extensions.

\subsubsection{Logging and Log Collection Module}

In the \textit{API Rewriting} section, we discuss using the \code{\seqsplit{Object.defineProperty}} function to modify existing properties. Logging functionality can also be injected into the process. For example, when modifying the value of \code{\seqsplit{navigator.userAgent}}, we can inject logging logic directly into the getter function:

\begin{lstlisting}
const originalValue = window.navigator.userAgent

Object.defineProperty(window.navigator, 'userAgent', {
  get() {
    logCollector.sendLog('userAgent') // send logs to the log collector
    return 'Forged UserAgent' // return a forged value here
  },
  configurable: true
})
\end{lstlisting}

The LogCollector is a class that collects logs and sends them when the number of logs reaches a limit or timeout. It is used to combine multiple logs to avoid listening misses when multiple custom events are sent at the same time.

It is important to note that the log module runs asynchronously. Since \code{Object.defineProperty} does not support asynchronous functions and real-time log retrieval is not necessary, this design ensures that the process does not prevent subsequent code from executing. This is achieved by sending a custom event with the log data, which the extension environment listens for and writes to \code{chrome.storage.local}. For more details, see the \textit{Data Transfer Process} section.

\subsubsection{Monitoring}

Using the log data collected in the previous step, the proposed solution enables users to visualize and analyze the log through interactive charts and detailed lists. This provides valuable insights, such as the frequency of digital fingerprint access attempts during a single request and the specific attributes being accessed. The log data are formatted and passed to the \code{BarChart} component of the \code{Recharts} library for visualization. In addition, a detailed list shows all URLs and their associated fingerprinting activity.


\section{Implementation}

\begin{figure*}[t]
\centering
\includegraphics[width=1\hsize]{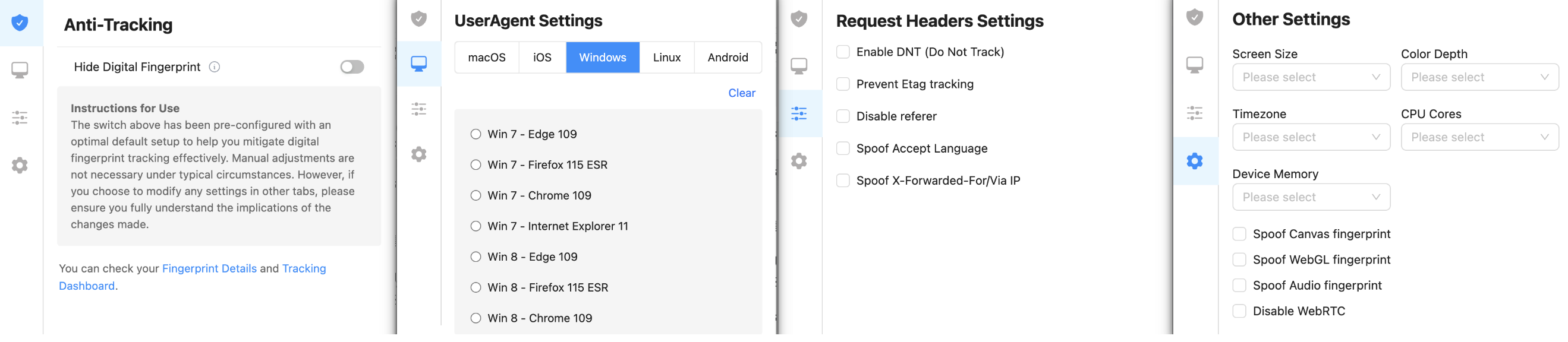}
\caption{Screenshots of our Chrome extension showing different menu options.}
\label{fig:menu}
\end{figure*} 

\begin{figure*}[t]
\centering
\includegraphics[width=0.9\hsize]{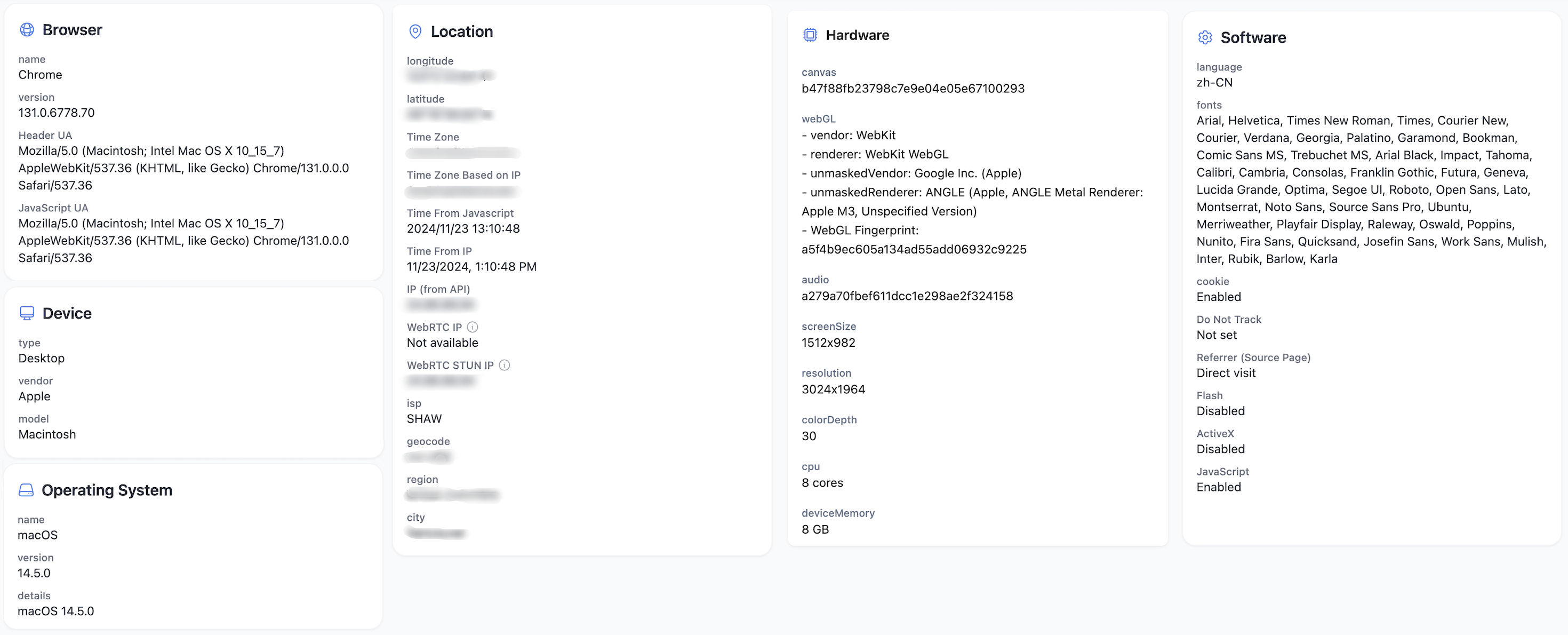}
\caption{A screenshot of fingerprint information.}
\label{fig:FingerprintInfo}
\end{figure*}

\subsection{Chrome Extension UI}

The proposed solution has a pop-up menu with four menu tabs.

\subsubsection{Front page}
On the front page, a switch allows users to get a random spoofed browser fingerprint. In addition, fingerprint details and tracking dashboards are available through the links at the bottom of the home page section, as shown in Figure \ref{fig:menu}.

\subsubsection{UserAgent Settings}
The extension provides five operating systems tabs, including three client systems, Windows, Linux, and macOS, and two mobile systems, Android and iOS. In each tab, multiple options are provided for combinations of system version and browser version. When users turn on the one-click defense on the homepage, the extension will randomly select an option in this part for simulation. They are also free to modify the option.

\subsubsection{Request Headers Settings}
The extension provides multiple options for header generation, including 'Do Not Track' flag, 'prevent eTag Tracking', 'Referer', 'accept language', and 'X-Forwarded' settings. When users turn on one-click protection on the homepage, the extension will enable some values in this section where they can set the options they need.

\subsubsection{Other Settings}
The extension provides some hardware simulation options and other fingerprint resistance settings in this section. Hardware-related settings include screen size, CPU cores, and device memory. Fingerprint resistance includes Canvas fingerprint, WebGL fingerprint, and audio fingerprint obfuscation. It also supports changes in time zones and WebRTC. When the user turns on a one-click defense on the homepage, the extension will be enabled in this section and randomly generate some values. They can also set the options they need here.






\subsection{Internal Webpage UI}

\subsubsection{Fingerprint Details}

The Fingerprint Details page displays the information in such modules: browser, device, operating system, location, hardware, software, and extensions. Users can use this information page to check whether the settings they made through the settings panel are effective. Figure \ref{fig:FingerprintInfo} shows an example of browser fingerprint details.

In the Browser section, the extension obtains and displays the \code{user-agent} information, browser, and version information of the user in the current tab. When users update the configuration in the current tab, the information reflects the modified disguise.

In the Device section, Operating System section, and Location section, the extension obtains and displays the user's device information, operating system information, and timezone information in the current tab and uses \code{ua-parser-js} to process and extract the information. Similarly, when the user modifies the relevant configuration in the current tab, this part of the information will display the modified disguised information.

In the Hardware section, the extension obtains and displays the user's hardware-related information in the current tab, such as screen size, color depth, CPU, and memory. The user can also modify these details through our extension. 

The extension provides a unique canvas fingerprint based on the \code{Canvas} rendering results for the user. This information is generated by drawing text and graphics on the \code{Canvas}, extracting the canvas data as Base64 encoding, and using MD5 hashing to process the \code{Canvas} data to obtain the unique \code{Canvas} fingerprint information.

The extension provides a \code{WebGL} fingerprint for the user. It uses \code{WebGL}'s parameter interface to extract the browser's \code{WebGL} vendor and renderer information, such as \code{gl.VENDOR} and \code{gl.RENDERER}. If the browser supports the \code{WEBGL\_debug\_renderer\_info} extension, more detailed GPU information (\code{unmaskedVendor} and \code{\seqsplit{unmaskedRenderer}}) can also be obtained. The extension renders fixed graphic content (such as a green rectangle) in the \code{WebGL} environment. It reads the pixel data of the rendering result (\code{gl.readPixels}) and converts the data into a unique fingerprint string through MD5 hashing.

The extension provides an audio fingerprint for the user. The generation method roughly creates an audio context, generates a fixed-frequency triangle wave signal, and uses \code{AnalyserNode} to extract its spectrum data. The extracted data is converted into a string, and a unique fingerprint is generated using MD5 hashing.

The information in the Software section can be modified using the Request Headers Settings options. The extension detects available font types by measuring text height, browser function enumeration, and \code{Canvas} detection. When the user turns on the function of disabling \code{WebRTC}, the location-related will not be displayed because this information is currently affected by the \code{WebRTC}.

\subsubsection{Tracking Dashboard}

On the Tracking Dashboard page, the user can view the log information of digital fingerprint-related indicators obtained by each web page when accessing the web page through the current browser.

\subsection{Performance Analysis}

To compare page loading performance and browser resource consumption before and after using the extension, the browser version tested is Chrome 131.0.6778.70 (arm64). We selected three websites for testing based on content complexity: Wikipedia (low-complexity), Instagram (medium-complexity), and YouTube (high-complexity). Table \ref{tab:performance_metrics} shows our test results. Each test result is based on an average of 50 tests. We can see no noticeable difference in the user's web browsing experience and actual browser resource usage before and after the extension is enabled.

\begin{table}[t]
\centering
\caption{Performance Metrics for Different Platforms}
\label{tab:performance_metrics}
{\footnotesize
\begin{tabular}{@{}llcc@{}}
\toprule
\textbf{Complexity} & \textbf{Platform} & \textbf{Get Configuration (ms)} & \textbf{API Rewrite (ms)} \\ \midrule
High & YouTube           & 37                              & 0.4                        \\
Medium & Instagram         & 36                              & 0.3                        \\
Low & Wikipedia         & 33                              & 0.2                        \\ \bottomrule
\end{tabular}
}
\end{table}

For storage, the impact of extensions on memory is mainly that the code injected by the extensions may affect the memory usage of the page\cite{heapSnapshots}\cite{chromeStorageMv2}. Table \ref{tab:memory_usage} is the average of the memory usage and total JS heap size of each website monitored every 10 seconds for 10 minutes with or without extensions enabled. According to the data in Table \ref{tab:memory_usage}, whether the extension is enabled or not has no significant effect on page memory usage.

\begin{table}[t]
\centering
\caption{Comparison of Total JS Heap Size and Memory Usage.}
\label{tab:memory_usage}
{\footnotesize
\begin{tabular}{@{}lcc@{}}
\toprule
\textbf{Test Case}              & \textbf{Total JS Heap Size} & \textbf{Memory Usage} \\ \midrule
Wikipedia (Disabling extension)    & 22.6 MB                    & 180 MB                \\
Wikipedia (Enabling extension)     & 23.7 MB                    & 181 MB                \\
Instagram (Disabling extension)    & 54.8 MB                    & 310 MB                \\
Instagram (Enabling extension)     & 54.5 MB                    & 310.7 MB                \\
YouTube (Disabling extension)      & 100 MB                     & 378 MB            \\
YouTube (Enabling extension)       & 97 MB                      & 370 MB                \\ \bottomrule
\end{tabular}
}
\end{table}

\section{Conclusion}

In this project, we developed a Chrome extension to detect, analyze, and customize browser fingerprints, addressing the increasing concern over online privacy and evolving user tracking methods. Our solution detects and logs browser fingerprints and provides users with tools to randomize and customize these fingerprints, adding a layer of protection against tracking mechanisms. Overall, this project demonstrates how browser extension technologies can be used to protect user privacy in a digital landscape where tracking and identification methods continue to evolve.

\bibliographystyle{ACM-Reference-Format}
\bibliography{refs}







\end{document}